\documentclass[12 pt]{article}

\usepackage{graphicx}

\begin{document}

\begin{center}

{In: ``Nonextensive Entropy - Interdisciplinary Applications'', edited by M. Gell-Mann and C. Tsallis,
New York Oxford University Press, 2003.}

\bigskip

{\LARGE Power-Law Persistence in the Atmosphere: \\ Analysis and Applications}
\bigskip

Armin Bunde$^a$, Jan Eichner$^a$, Rathinaswamy Govindan$^{a,b}$, Shlomo Havlin$^b$, \\
Eva Koscielny-Bunde$^{a,c}$, Diego Rybski$^{a,b}$ and Dmitry Vjushin$^b$
\bigskip

{\it $^a$Institut f\"ur Theoretische Physik III, Universit\"at Giessen, D-35392 Giessen, Germany}

{\it $^b$Minerva Center and Department of Physics, Bar-Ilan University, Israel}

{\it $^c$Potsdam Institute for Climate Research, D-14412 Potsdam, Germany}

(submitted: 20 December 2002, revised: 31 January 2003)
\end{center}

\begin{abstract}
We review recent results on the appearance of long-term persistence 
in climatic records and their relevance for the evaluation of global climate models and rare events.
The persistence can be characterized, for example, by the correlation $C(s)$ of temperature
variations separated by $s$ days. We show that, contrary to previous expectations, 
$C(s)$ decays for large $s$ as a power law, $C(s)\sim s^{-\gamma}$. For continental stations, 
the exponent $\gamma$ is always close to 0.7,
while for stations on islands $\gamma \cong 0.4$. In contrast to the
temperature fluctuations, the fluctuations of the rainfall usually cannot be characterized by 
long-term power-law correlations but rather by pronounced short-term correlations.
The universal persistence law for the temperature fluctuations on continental stations represents 
an ideal (and uncomfortable) test-bed for the state of-the-art global
climate models and allows us to evaluate their performance. In addition, the presence of long-term 
correlations leads to a novel approach for evaluating the statistics of rare events.
\end{abstract}

\section*{\large{1. INTRODUCTION}}
The persistence of  weather states on short terms is a well-known phenomenon:
a warm day is more likely to be followed by a warm day than by a cold
day and vice versa. The trivial forecast that the weather of tomorrow is the same as the weather 
of today was, in previous times, often used as a "minimum skill" forecast for assessing
the usefulness of short-term weather forecasts. The typical time scale for weather changes is 
about one week, a time period which corresponds to the average duration of so-called
``general weather regimes" or ``Grosswetterlagen'', so this type of short-term persistence
usually stops after about one week. On larger scales, other types of persistence
occur, one of them is  related to circulation patterns associated with blocking \cite{B8}. 
A blocking situation occurs when a very stable high-pressure system is established over a particular
region and remains in place for several weeks. As a result, the weather in the region of the high
remains fairly persistent throughout this period. Furthermore, transient low-pressure
systems are deflected around the blocking high so that the region downstream of the
high experiences a larger than usual number of storms. On even longer terms, a source for weather 
persistence might be slowly varying external (boundary), forcing such as sea surface temperatures
and anomaly patterns for example. On the scale of months to seasons, one of the most pronounced phenomenon is 
the El Ni$\tilde{n}$o Southern Oscillation (ENSO) event, which occurs every three to five years and which strongly affects 
the weather over the tropical Pacific as well as over North America \cite{B9}.

The question is, {\it how } the persistence, that might be generated by very different mechanisms 
on different time scales, decays with time $s$. The answer to this question is not simple. 
Correlations, and in particular long-term correlations, can be masked by trends
that are generated, for instance, by the well-known urban warming phenomenon. Even uncorrelated data in the 
presence of long-term trends may look like correlated data, and, on the other
hand, long-term correlated data may look like uncorrelated data influenced by a trend.

Therefore, in order to distinguish between trends and correlations, one needs methods
that can systematically eliminate trends. Those methods are available now: both
wavelet techniques (WT)$^1$  and detrended fluctuation analysis (DFA)$^2$ can systematically eliminate trends in
the data and thus reveal intrinsic dynamical properties such as distributions, scaling and long-range 
correlations that are often masked by nonstationarities. 

In recent studies \cite{Eichner, EVA, EVA1, Rybski} we have used DFA and WT to study temperature and precipitation 
correlations in different climatic zones on the globe.
The results indicate that the temperature variations are long-range power-law correlated above some 
crossover time that is of the order of 10 days. Above 10 d,
the persistence, characterized by the autocorrelation $C(s)$
of temperature variations separated by $s$ days, decays as 
$$C(s)\sim s^{-\gamma}, \eqno(1)$$
where, most interestingly, the exponent $\gamma$ has roughly the same value $\gamma\cong 0.7$ for
all continental records. For small islands  the correlations are more pronounced, with
$\gamma$ around 0.4. This value is close to the value obtained recently for correlations 
of sea-surface temperatures \cite{Roberto}. In marked contrast, for most stations the precipitation
records do not show indications of long-range temporal correlations on scales above 6 months. 
Our results are supported by independent analysis by several groups \cite{P2, P1, T1}.

The fact that the correlation exponent varies only very little for the continental atmospheric temperatures, 
presents an ideal test-bed for the performance of the global climate models, as we will show below.
We present an analysis of the two standard scenarios (greenhouse gas 
forcing only and greenhouse gas plus aerosols forcing) together with the analysis of a control run. 
Our analysis  points to clear deficiencies
of the models.  For further discussions we refer to Govindan et al. \cite{GOVIN1, GOVIN}.
Finally, we review a recent approach to determine the statistics of rare events in the
presence of long-term correlations. 

The chapter is organized in five sections. In Section 2, we describe one of the detrending analysis methods, 
the detrended fluctuation analysis (DFA). In Section 3, we review the application of this method to both 
atmospheric temperature and  precipitation records. In Section 4, we describe how the "universal" 
persistence law for the atmospheric temperature fluctuations on continental stations can be used to test 
the three scenarios of the state-of-the-art climate models. In Section 5, finally, we describe how the common 
extreme value statistics is modified in the presence of long-term correlations.

\section{THE METHODS OF ANALYSIS}

Consider, for example, a record $T_i$, where the index $i$ counts the days in the record, $i=1$,2,...,$N$.
This record $T_i$ may represent the maximum daily temperature  or the daily amount of precipitation, measured 
at a certain meteorological station. For eliminating the periodic seasonal trends, we concentrate on the 
departures of the $T_i$, $\Delta T_i=T_i - \overline T_i$, from their mean daily
value $\overline T_i$ for each calendar date $i$, say 1st of
April, which has been obtained by averaging over all years in the record.

Quantitatively, correlations between two $\Delta T_i$ values separated by
$n$ days are defined by the (auto)-correlation function
$$C(n) \equiv \langle \Delta T_i \Delta T_{i+n}\rangle={1\over N-n}
\sum_{i=1}^{N-n}\,,\,\Delta T_i\Delta T_{i+n} \eqno(2).$$
If the $\Delta T_i$ are uncorrelated, $C(n)$ is zero for
$n$ positive. If correlations exist up to a certain number of days
$n_\times$, the correlation function will be positive up to $n_\times$
and vanish  above $n_\times$. A direct calculation of $C(n)$ is
hindered by the level of noise present in the finite
records, and by possible nonstationarities in the data.

To reduce the noise we do not calculate $C(n)$ directly, but instead study the ``profile''
$$Y_m=\sum_{i=1}^m \Delta T_i.  \eqno(3).$$
We can consider the profile $Y_m$ as the position of a random walker on a linear chain after
$m$ steps. The random walker starts at the origin and performs, in the $i$th step, a jump of length
$\Delta T_i$ to the right if  $\Delta T_i$ is positive, and to the left if  $\Delta T_i$ is
negative. The fluctuations ${F^2(s)}$ of the profile,
in a given time window of size $s$, are related to the correlation function $C(s)$.
For the relevant case (1) of long-term power-law correlations, \quad $C(s)\sim s^{-\gamma},\quad 0<\gamma<1,$
the mean-square fluctuations $\overline{F^2(s)}$, obtained by averaging over many time windows of size $s$ 
(see below) asymptotically increase by a power law \cite{BU1},$$
\overline{F^2(s)}\sim s^{2\alpha}, \quad\alpha=1-\gamma/2.\eqno(3)$$
For uncorrelated data (as well as for  correlations
decaying faster than $1/s$), we have $\alpha=1/2$.

For the analysis of the fluctuations, we employ a hierarchy of methods that differ in the way the
fluctuations are measured and possible trends are eliminated (for a detailed description of the methods
we refer to Kantelhardt et al. \cite{KANT}).

1.) In the simplest type of fluctuation analysis (FA)  (where trends are not going to be eliminated),
we determine the difference of the profile at both ends of each window. The square of this difference
represents the square of the fluctuations in each window.

2.) In the {\it first-order} detrended fluctuation analysis (DFA1), we
determine in each window the best linear fit of the profile.
The variance of the profile from this straight line
represents the square of the fluctuations in each window.

3.) In general, in the $n$th order DFA  (DFAn) we determine in each window
the best $n$th order polynomial fit of the profile. The
variance of the profile from these best $n$th-order polynomials
represents the square of the fluctuations in each window.

By definition, FA does not eliminate trends similar to the Hurst method
and the conventional power spectral methods \cite{FEDER}. In contrast,
DFAn eliminates trends of order $n$ in the profile and $n-1$
in the original time series. Thus, from the comparison of
fluctuation functions $F(s)$ obtained from different methods, one can
learn about long-term correlations and types of trends, which cannot
be achieved by the conventional techniques.

\section*{\large{2. ANALYSIS OF TEMPERATURE AND PRECIPITATION RECORDS}}

\noindent
Figure 1  shows the results of the FA and DFA analysis  of the maximum daily temperatures
$T_i$ of the following weather stations (the length of the records is written within the
parentheses): (a) Cheyenne (USA, 123 y), (b) Edinburgh (UK, 102 y), (c) Campbell Island 
(New Zealand, 57 y), and  (d) Sonnblick (Austria, 108 y). 
The results are typical for a large number of records that we have 
analyzed so far (see Eichner et al.\,\cite{Eichner} and Koscielny-Bunde et al.\,\cite{EVA, EVA1}). 
Cheyenne has a continental climate, Edinburgh is on a coastline, Campbell Island is a small 
island in the Pacific Ocean, and the weather station of Sonnblick is on top of a mountain.

\begin{figure}
\centering
\includegraphics[height=4in]{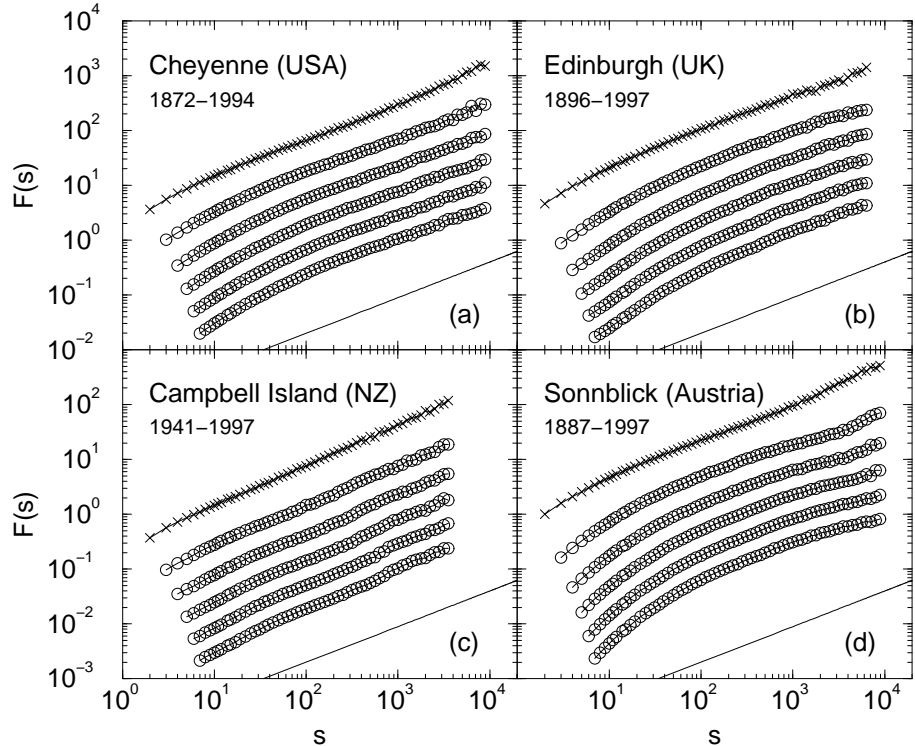}
\caption{Analysis of daily temperature records of four representative weather
stations. The four figures show the fluctuation functions obtained by FA, DFA1, DFA2, DFA3, DFA4, 
and DFA5 (from top to bottom) for the four sets of data. The scale of the fluctuation functions is
arbitrary. In each panel, a line with slope 0.65 is shown as a guide to the eye.}
\end{figure}

In the log-log plots, all curves are (except at small $s$ values) approximately straight lines. For both
the stations inside the continents and along coastlines, the slope is $\alpha \cong 0.65$. There exists
a natural crossover (above the DFA crossover) that can be best estimated from FA and DFA1. As can be verified easily, 
the crossover occurs roughly at $t_c=10d$, which is the order of magnitude 
for a typical Grosswetterlage. Above $t_c$,  there exists long-range persistence expressed  by the 
power-law decay of the correlation function with an exponent $\gamma = 2 -2 \alpha \cong 0.7$. 
These results are representative for the large number of records we have analyzed. They indicate that the exponent 
is "universal" , i.e., does not depend on the location and the
climatic zone of the weather station. Below $t_c$, the fluctuation functions do not show universal 
behavior and reflect the different climatic zones.

However, there are exceptions from the universal behavior, and these occur for locations
on small islands and on top of large mountains. In the first case, the exponent can be considerably 
larger, $\alpha \cong 0.8$, corresponding to $\gamma \cong 0.4$. In the second case, on top of a mountain,
the exponent can be smaller, $\alpha \cong 0.58$, corresponding to $\gamma \cong 0.84$.

Next we consider precipitation records.
Figure 2 shows the results of the FA and DFA analysis  of the daily precipitation
$P_i$ of the following weather stations (the length of the records is written within the
parentheses): (a) Cheyenne (USA, 117 y), (b) Edinburgh (UK, 102 y), (c) Campbell Island 
(New Zealand, 57 y), and (d) Sonnblick (Austria, 108 y).
The results are typical and represent a large number of records that we have analyzed 
so far \cite{Rybski}).

\begin{figure}
\centering
\includegraphics[height=4in]{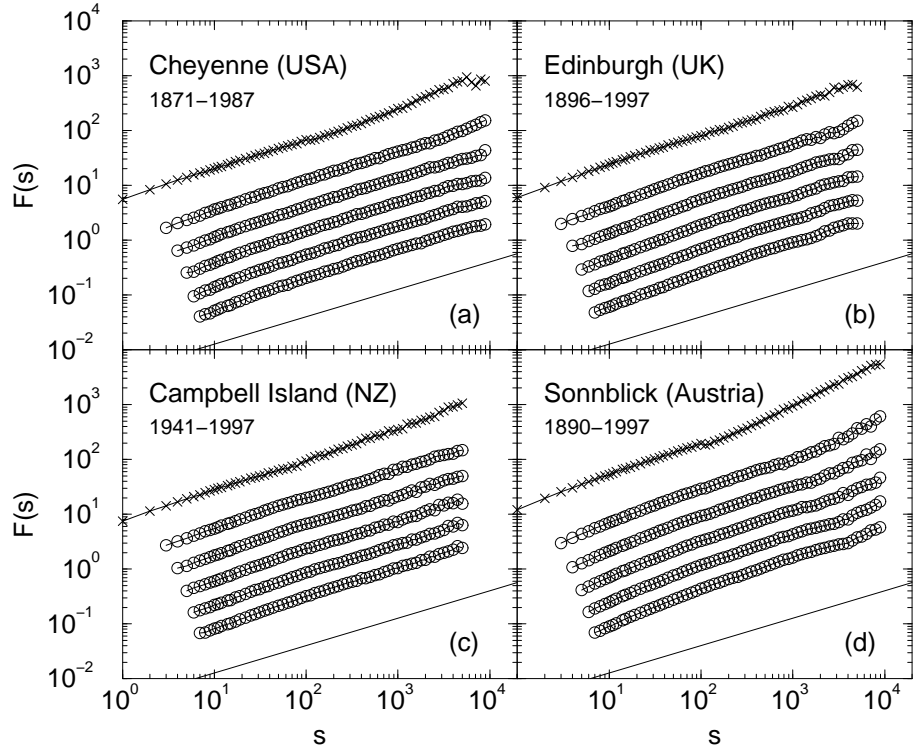}
\caption{Analysis of daily precipitation records of four representative weather stations. 
The four figures show the fluctuation functions obtained by FA, DFA1, DFA2, DFA3, DFA4, 
and DFA5 (from top to bottom) for the four sets of data. The scale of the fluctuation 
functions is arbitrary. In each panel, a line with slope 0.5 is shown as a guide to the eye. }
\end{figure}

In the log-log plots, all curves are (except at small $s$ values) approximately straight lines at large times,
with a slope close to 0.5. If there exist long-term correlations, then they are very small. Some exceptions
are again stations on top of a mountain, where the exponent might be around 0.6, but this happens only very
rarely. In most cases, the exponent is between 0.5 and 0.55, pointing to uncorrelated or weakly correlated
behavior at large time spans. Unlike the temperature records, the exponents actually do not depend on specific
climatic or geographic conditions.

\begin{figure}
\centering
\includegraphics[height=4in]{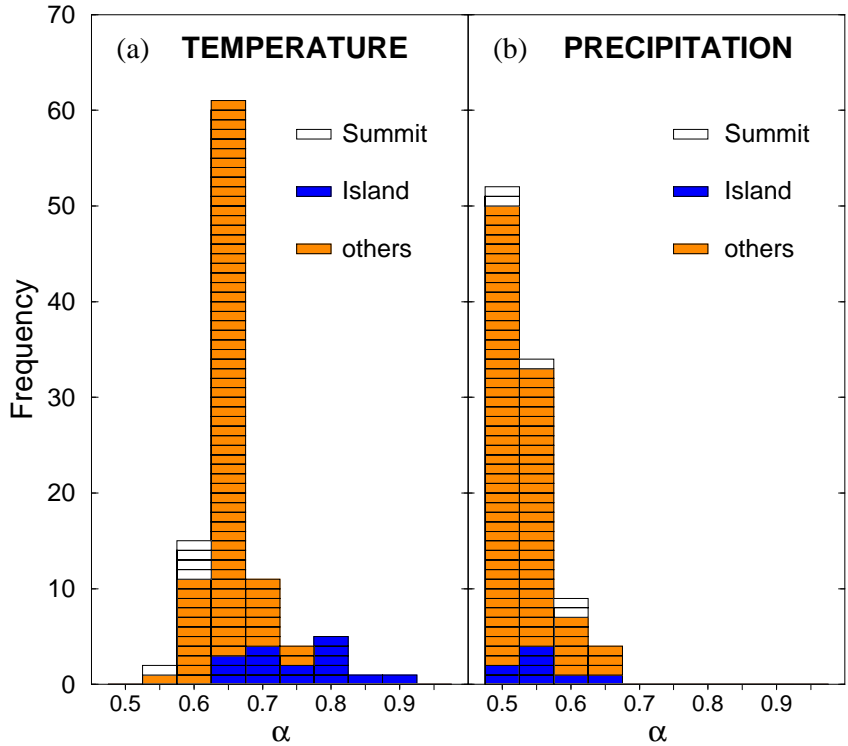}
\caption{Histograms of the values of the fluctuation exponents $\alpha$ (a) for daily temperature records and (b) for daily 
precipitation records. }
\end{figure}

Figure 3 summarizes the results for exponents $\alpha$ for (a) temperature records
and (b) precipitation records. Different climatological conditions are marked in the
histograms. First we concentrate on the temperature records (fig. 3a). One can see clearly that for 
stations that are neither on islands nor on summits, the average exponent is close 0.65, with
a variance of  0.03. For the islands (where only few records are available) the average
value of $\alpha$ is 0.78, with quite a large variance of 0.08.  The variance is large, since stations on 
larger islands, like Wrangelija, behave more like continental stations, with an exponent close to 0.65. 
For the precipitation records (fig. 3b), the average exponent $\alpha$ is close to
0.54, with a variance close to 0.05, and does not depend significantly on the climatic conditions around 
a weather station.

Since for the temperature records the exponent for continental and coastline stations does not depend on the location of the 
meteorological station and its local environment, the power-law behavior can serve as an ideal 
test for climate models where regional details cannot be incorporated and therefore regional phenomena 
like urban warming cannot be accounted for. The power-law behavior seems to be a global phenomenon and 
therefore should also show up in the simulated data of the global climate models (GCM).

\section*{\large{3. TEST OF GLOBAL CLIMATE MODELS}}

The state-of-the-art  climate models that are used to estimate future climate are coupled 
atmosphere-ocean general 
circulation models (AOGCMs) \cite{J3, INT}.
The models provide numerical solutions of the Navier Stokes 
equations devised for simulating meso-scale to large-scale atmospheric and oceanic 
dynamics. In addition to the explicitly resolved scales of motions, the models 
also contain parameterization schemes representing the so-called subgrid-scale 
processes, such as radiative transfer, turbulent mixing, boundary layer processes, 
cumulus convection, precipitation, and gravity wave drag. A radiative transfer scheme, 
for example, is necessary for simulating the role of various greenhouse gases such 
as CO$_2$ and the effect of aerosol particles.   
The differences among the models usually lie in the selection of the 
numerical methods employed, the choice of the spatial resolution$^3$, and the 
subgrid-scale parameters. 

Three scenarios  have been studied by the models, 
and the results are available, for four models, from the IPCC Data Distribution Center \cite{IPCC}.
The first scenario represents a control run where the CO$_2$ content is kept fixed.
In the second scenario, one considers only the effect of greenhouse gas forcing (GHG). 
The amount of greenhouse gases is taken from the observations  until 1990 and then 
increased at a rate of 1\%
per year. In the third scenario, the effect of aerosols (mainly sulfates)
in the atmosphere is taken into account. Only direct sulfate
forcing  is considered;  until 1990, the sulfate concentrations are taken from
historical measurements , and  are increased linearly afterwards. The effect of sulfates 
is to mitigate and partially offset the greenhouse gas warming.
Although this scenario represents an important step
towards comprehensive climate simulation, it introduces
new uncertainties - regarding the distributions of natural and
anthropogenic aerosols and, in particular, regarding indirect effects
on the radiation balance through cloud-cover modification etc. \cite{WGI}.

\begin{figure}
\centering
\includegraphics[height=7in]{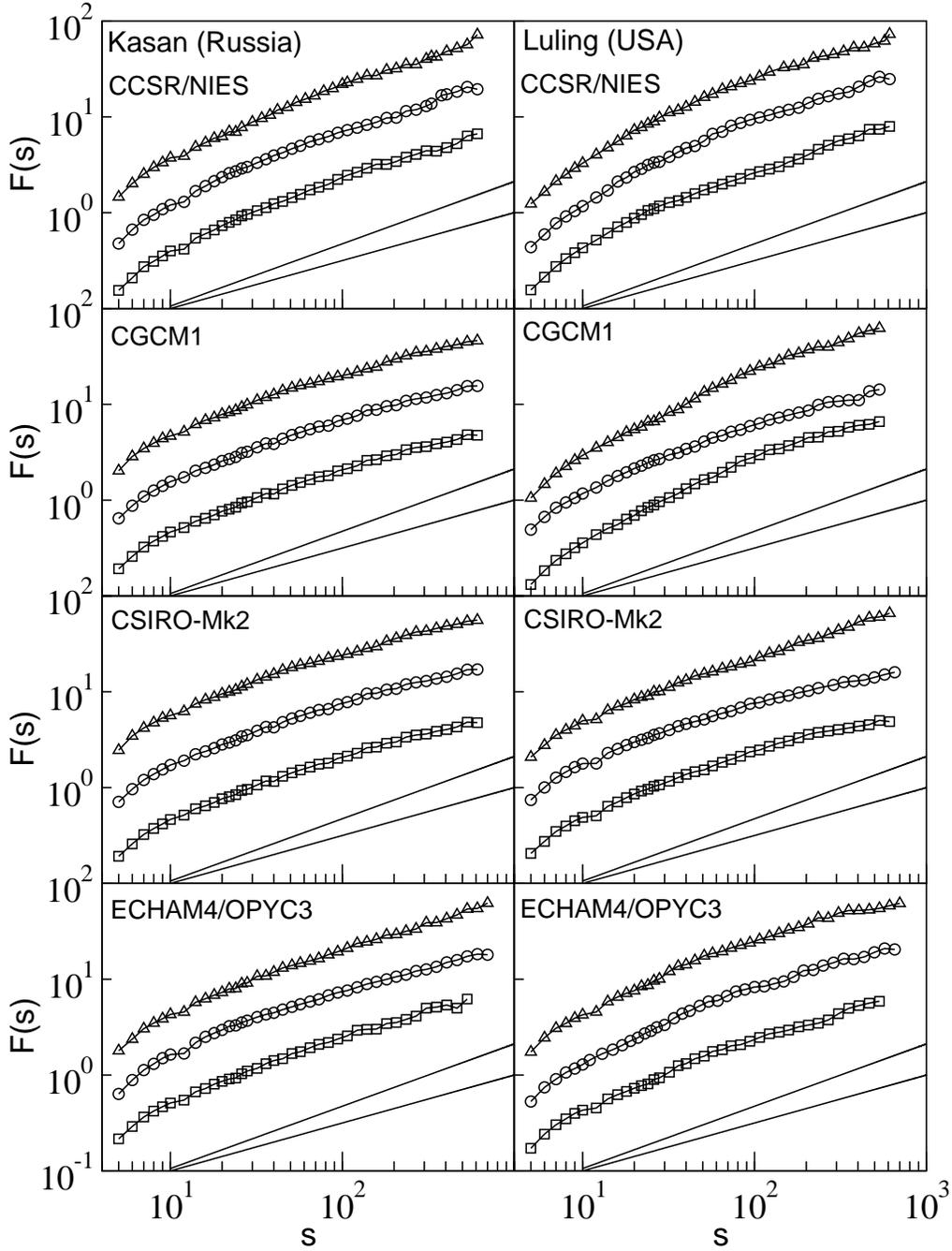}
\caption{Comparison of the scaling performance of the three scenarios: control run ($\triangle$), greenhouse gas
forcing only ($\circ$) and greenhouse gas plus aerosols ($\diamond$). All curves are obtained by applying 
DFA3 to  the  monthly mean of the daily maximum temperatures generated by the four AOGCMs.  The lines with slopes 
0.65 and 0.5 are shown  as a guide to the eye. For details of the records, we refer to \cite{IPCC}.}
\end{figure}

For the test, we consider the monthly temperature records  from  those four AOGCMs. 
Data for these three scenarios are available from the Internet: CSIRO-Mk2 (Melbourne), 
CCSR/NIES (Tokyo), ECHAM4/OPYC3 (Hamburg), and CGCM1 (Victoria, Canada). We extracted
the data for six representative sites around the globe  (Prague, Kasan, Seoul,
Luling [Texas], Vancouver, and Melbourne). For each model and each of the three scenarios, 
we selected the temperature records of the four grid points closest to each site, and bilinearly 
interpolated the data to the location of the site.
Figure 4 shows representative results of the fluctuation functions, calculated using DFA3,
for two sites (Kasan [Russia] and Luling [Texas]) for the four models and the three scenarios.
As seen in figure 4 most of the DFA curves approach the slope of 0.5. However, the control runs seem to show a 
somewhat better performance, i.e., many of them have a slope close to 0.65 (e.g., Luling (CSIRO-Mk2)),
and the greenhouse gas only scenario show the worst performance.
The actual long-term exponents $\alpha$ for the three scenarios of the 4 models
for the 6 cities are summarized in figure 5(a)-(c). Each histogram consists of 24 blocks and every block is 
specified by the model and the city.

\begin{figure}
\includegraphics[width=1.0\textwidth]{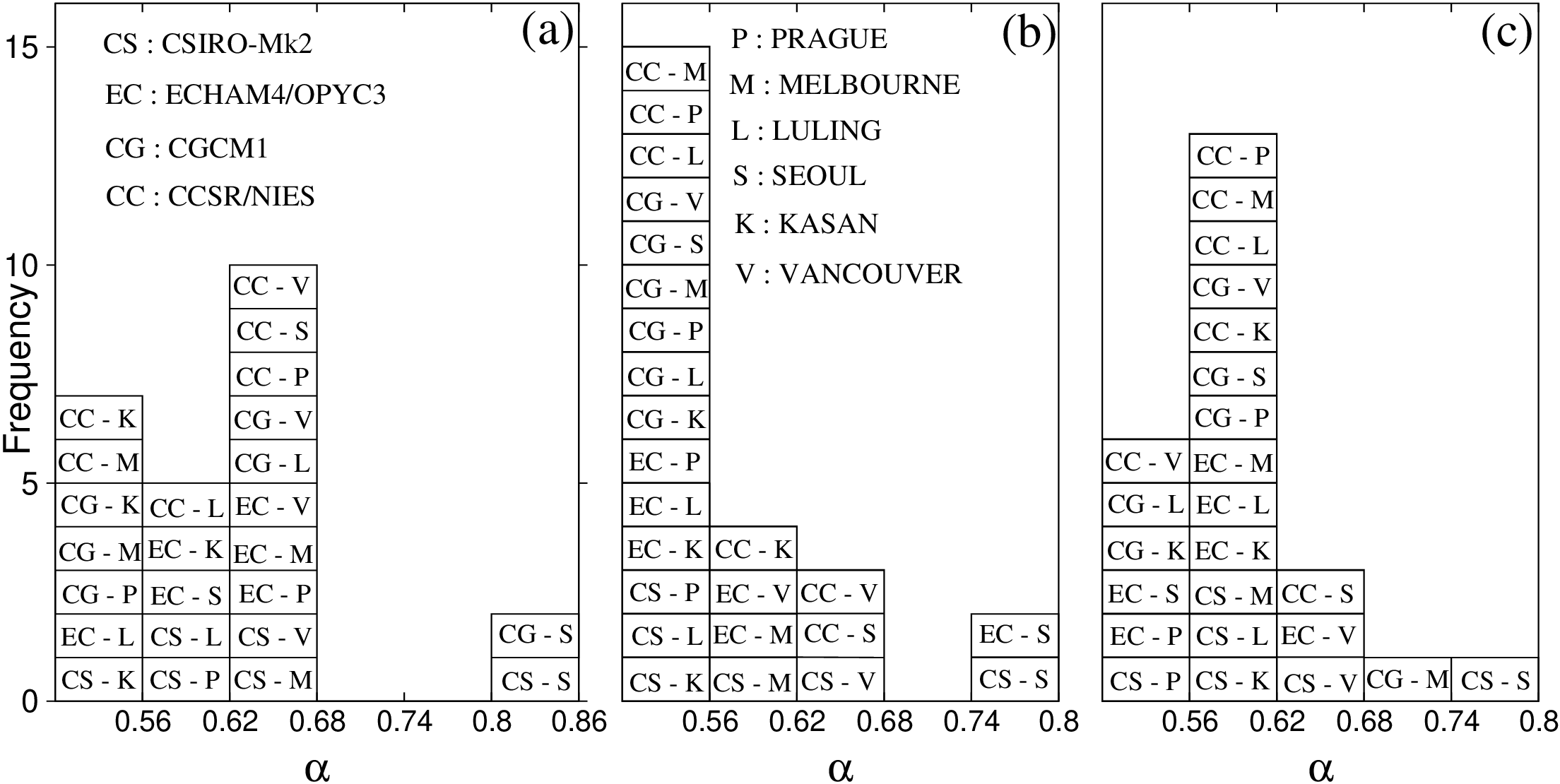}

\caption{Histograms of the values of the fluctuation exponent ($\alpha$) obtained from the simulations of the four 
AOGCMs (listed in (a)), for six sites (listed in (b)). The three panels are for the three scenarios: (a) control run, 
(b) greenhouse gas forcing only,  and (c) greenhouse gas plus aerosol forcing.
The entries in each box represent "Model - Site".}
\end{figure}

For the control run (fig. 5(a)) there is a peak at $\alpha \cong 0.65$ but more than half of the exponents
are below $\alpha \cong 0.62$. 
For the greenhouse gas only scenario (fig. 5(b)), the histogram shows a pronounced maximum at $\alpha=0.5$. 
For best performance, all models should have exponents $\alpha$  close to 0.65, corresponding to a peak of
height 24 in the window between 0.62 and 0.68. Actually, more than half of the exponents are close to 0.5, 
while only 3 exponents are in the proper window between 0.62 and 0.68.
Figure 5(c) shows the histogram for the greenhouse gas  plus aerosol scenario, where, in addition to the greenhouse gas forcing, 
also the effects of aerosols are taken into account. For this case, there is a pronounced maximum in 
the $\alpha$ window between 0.56 and 0.62 (more than half of the exponents are in this window), while again
only 3 exponents are in the proper range between 0.62 and 0.68. This shows that
although the greenhouse gas  plus aerosol scenario is also far from reproducing the scaling behavior of the real data, 
its overall performance is better than the performance of the greenhouse gas  scenario.
The best performance is observed for the control run, which points to remarkable deficiencies in the way the forcings are 
introduced into the models.

\section*{\large{4. EXTREME VALUE STATISTICS}}

The long-term correlations in a record $\lbrace{T_i\rbrace}$ effect strongly the statistics of the extreme values in the record, as has been shown
recently in [2]. The central quantity in Extreme Value Statistics (EVS) is the return time $r_q$ between two events of size greater or equal to a 
certain threshold $q$. The basic assumption in conventional EVS is, that the events are uncorrelated (at least when the time lag between them is 
sufficiently large). In this case, one can obtain the mean return time $R_q$ simply from the probability $W_q$ that an event greater or equal
to $q$ occurs, $R_q\,=\,1 / W_q$. Since the events are uncorrelated, also the return intervals are uncorrelated and follow the Poisson-statistics; 
i.e. their distribution function $P_q(r)$ is a simple exponential, $P_q(r)\,\sim\,\exp(-r/R_q)$.\\
\indent For long-term correlations, it has been shown in [2] that the distribution function $P_q(r)$ changes into a {\it{stretched}} exponential 
function,
$$P_q(r)\,\sim\,{\rm{\exp}}[-{\rm{const}}(r/R_q)^\gamma)], \eqno(4)$$
for $\gamma$ between zero and one. For $\gamma$ above one, in the case of short-term correlations, $P_q$ reduces to the Poisson distribution.

\begin{figure}
\centering
\includegraphics[height=3in]{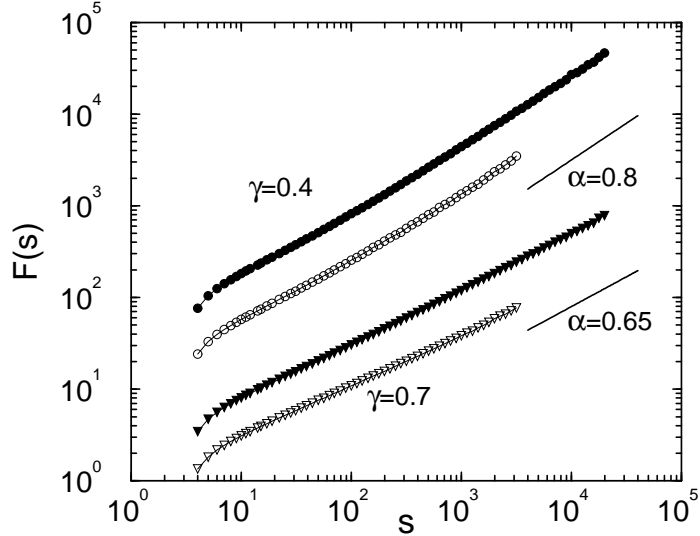}
\caption{Fluctuation functions $F(s)$ for the record of return intervals obtained from two artificial long-term correlated records $\lbrace{T_i}\rbrace$ with $\gamma\,=\,0.4$
(upper curves) and $\gamma\,=\,0.7$ (lower curves). The distribution of the $T_i$ values has been chosen as Gaussian, with zero mean and variance one. For the return intervals, 
the thresholds $q\,=\,1.5$ and $2.5$ have been considered. The straight lines in the figure have slopes $\alpha\,=\,1\,-\,\gamma/2$, suggesting that the return intervals are long-term
correlated in the same way as the $T_i$}
\end{figure}

\indent In addition, the return intervals become long-term correlated, with an exponent that is approximately identical to $\gamma$ [2]. This is seen in Fig. 6, where the 
fluctuation functions $F(s)$ of the return intervals (obtained by DFA2) are shown for two artificial long-term correlated records with $\gamma\,=\,0.4$ and 0.7. Two values 
of thresholds $q\,=\,1.5$ and $2.5$ have been considered for each value of $\gamma$. The distribution of $T_i$ values has been chosen as Gaussian, with zero mean and variance 
one. In the double logarithmic plot, all the curves approach straight lines with slopes $\alpha\,=\,1\,-\,\gamma/2$, suggesting that the return intervals are long-term correlated
in the same way as the $T_i$.\\
\indent As a consequence, small return intervals are more likely to be followed by small intervals and large intervals are more likely to be followed by large intervals. Accordingly,
for long-term correlated records it is more likely than for uncorrelated records that a sequence of large return times is followed by a sequence of short return times.\\
\indent This fact may be relevant for the occurence of floods. It is well known that river flows are long-term correlated with exponents $\gamma$ between 0.3 and 0.9, in most cases
close to 0.4. In the last decades, the frequency of large floods in Europe has increased. It is possible, that this increase is due to global warming, but it is also possible that it has
been triggered by the long-term correlations.
 
\section*{\normalsize{Acknowledgments}}
We are grateful to Prof. H.J. Schellnhuber, Dr. J.W. Kantelhardt, and Prof. S. Brenner for very useful discussions. We like to acknowledge financial support by the Deutsche Forschungsgemeinschaft 
and the Israel Science Foundation.


\begin{thebibliography}{99}

\bibitem{ARNEODO} Arneodo, A., dAubentonCarafa, Y., Bacry, E., Graves, P.V., Muzy, J.F., Thermes, C.
``Wavelet based fractal analysis of DNA sequences''. Physica D 96 (1996): 291-320.

\bibitem{BUNDE} Bunde, A., Eichner, J., Havlin, S., Kantelhardt, J.W. ``The Effect of Long-Term 
Correlations on the statistics of Rare Events''. Physica A (2003), in press.

\bibitem{BU1} Bunde, A. and Havlin, S. (eds.). Fractals in Science. New York: Springer, 1995.

\bibitem{HEART3} Bunde, A., Havlin, S., Kantelhardt, J.W., Penzel, T., Peter, J.H., Voigt, K.  
``Correlated and uncorrelated regions in heart-rate fluctuations during sleep''.
Phys. Rev. Lett. 85 (2000): 3736-3739.

\bibitem{B8} Charney, J.G., and Devore, J.G., J. Atmos. Sci. 36 (1979): 1205.

\bibitem{Eichner}  Eichner, J.,  Bunde,  A., Havlin,  S., Koscielny-Bunde, E., Schellnhuber,
H.J. ``Power-law persistence and trends in the atmosphere: A detailed study of long temperature
records''. Phys. Rev. E (2003) submitted.

\bibitem{FEDER} Feder, J. Fractals. New York: Plenum, 1989.

\bibitem{GOVIN1}
Govindan, R. B., Vjushin, D., Brenner, S., Bunde, A., Havlin, S., Schellnhuber, H.-J.
``Global climate models violate scaling of the observed atmospheric variability''.
Phys. Rev. Lett. 89  (2002): 028501.

\bibitem{GOVIN}
Govindan, R. B., Vjushin, D., Brenner, S., Bunde, A., Havlin S., H.-J. Schellnhuber H.-J.
``Long-range correlations and trends in global climate models: Comparison with real data''.
Physica A 294 (2001): 239; Vjushin, D., Govindan, R. B., Brenner, S., Bunde, A., Havlin, S., 
Schellnhuber, H.-J. ``Lack of scaling in global climate models''. J. Phys. C 14 (2002): 2275.

\bibitem{J3} Hasselmann, K. ``Multi-pattern fingerprint method for detection and attribution of climate change'',
Multi-fingerprint detection and attribution analysis of greenhouse gas, greenhouse gas-plus-aerosol and solar 
forced climate change''. Climate Dynamics 13 (1997): 601-634 and references therein.

\bibitem{WGI} Houghton, J.T. (editor). ``Climate Change 2001: The Scientific Basis, Contribution of Working
Group I to the Third Assessment Report of the Intergovernmental Panel on Climate Change (IPCC)''.  
Cambridge: Cambridge University Press, 2001.

\bibitem{INT} Intergovernmental Panel on Climate Change. {\it{The Regional Impacts of Climate Change. An Assessment
of Vulnerability}}, edited by R. T. Watson, M. C. Zinyowera, and R. H. Moss. Cambridge: Cambridge University Press,
1998.

\bibitem{IPCC} Data Distribution Center.
http://ipcc-ddc.cru.uea.ac.uk/dkrz/dkrz\_index.html

\bibitem{KANT} Kantelhardt, J. W., Koscielny-Bunde, E., Rego, H. A., Havlin, S., Bunde, A.
``Detecting long-range correlations with detrended fluctuation analysis''.
Physica A 295 (2001): 441-454.

\bibitem{EVA1}Koscielny-Bunde, E., Bunde, A., Havlin, S., Goldreich, Y. ``Analysis of daily temperature fluctuations''.
Physica A 231 (1996): 393-396.

\bibitem{EVA} Koscielny-Bunde, E., Bunde, A., Havlin, S., Roman, H. E.,
Goldreich, Y., Schellnhuber, H.-J. ``Indication of a universal persistence law governing atmospheric variability''. 
Phys. Rev. Lett. 81 (1998): 729-732.

\bibitem{Roberto}
Monetti, R. A., Havlin, S., and Bunde, A.  ``Long-term persistence in the sea surface temperature fluctuations''.
Physica A 320 (2003): 581-588.

\bibitem{P1}
Pelletier, J.D. ``Analysis and modeling of the natural variability of climate''.
J. Climate 10 (1997): 1331-1342.

\bibitem{P2}
Pelletier, J.D., Turcotte, D.L. ``Long-range persistence in climatological and hydrological time series: 
analysis, modeling and application to drought hazard assessment''. J. Hydrol. 203 (1997): 198-208.

\bibitem{DNA1}
Peng, C.-K., Buldyrev, S. V., Havlin, S., Simons, M., Stanley, H. E. , Goldberger, A. L.
``Mosaic Organization of DNA Nucleotides''. Phys. Rev. E 49 (1994): 1685-1689.

\bibitem{B9} Philander, S.G. ``El Nino, La Nina and the Southern Oscillation''. 
International Geophysics Series, Vol 46 (1990).

\bibitem{Rybski} Rybski, D., Bunde, A., Havlin, S., Schellnhuber, H.J. 
``Detrended fluctuation analysis of precipitation records: Scaling and multiscaling''. Preprint (2002).

\bibitem{T1}
Talkner, P., Weber, R.O. ``Power spectrum and detrended fluctuation analysis: Application to 
daily temperatures''. Phys. Rev. E 62 (2000): 150-160.

\bibitem{JUSH} Vjushin, D., R. B. Govindan, S. Brenner, A. Bunde, S. Havlin, and H.-J. Schellnhuber. 
``Lack of Scaling in Global Climate Models''. J. Phys. C 14 (2002): 2275.


\end{thebibliography}
\end{document}